\documentclass[review]{elsarticle}
\usepackage{epstopdf}
\usepackage{setspace}
\usepackage{bm}
\usepackage{latexsym,amssymb,rotating}
\usepackage[aboveskip=0pt,belowskip=0pt]{caption}
\usepackage{amsmath}
\usepackage{tabularx}
\usepackage{listings}
\usepackage{url}
\usepackage{subfig}

\def\4{\phantom{4}}

\usepackage{epsfig}
\usepackage{times}

\usepackage{psfrag}
\usepackage{longtable}

\begin{document}
\lstset{language=Fortran}

\title{$ms2$: A molecular simulation tool for thermodynamic properties, new version release}

\author[hlrs]{Colin W. Glass}
\author[ltd]{Steffen Reiser}
\author[upb]{G\'{a}bor Rutkai}
\author[ltd]{Stephan Deublein} 
\author[upb]{Andreas K\"{o}ster}
\author[upb]{Gabriela Guevara-Carrion}
\author[hlrs]{Amer Wafai}
\author[ltd]{Martin Horsch}
\author[hlrs]{Martin Bernreuther}
\author[upb]{Thorsten Windmann}
\author[ltd]{Hans Hasse}
\author[upb]{Jadran Vrabec \corref{cor1} }

\address[hlrs]{H\"{o}chstleistungsrechenzentrum Universit\"{a}t Stuttgart (HLRS), 70550 Stuttgart, Germany}
\address[ltd]{Lehrstuhl f\"{u}r Thermodynamik, Universit\"{a}t Kaiserslautern, 67653 Kaiserslautern, Germany}
\address[upb]{Lehrstuhl f\"{u}r Thermodynamik und Energietechnik, Universit\"{a}t Paderborn, 33098~Paderborn, Germany}

\cortext[cor1]{Corresponding author: Jadran Vrabec, Warburger Str. 100, 33098 Paderborn, Germany, Tel.: +49-5251/60-2421, Fax:
+49-5251/60-3522, Email: jadran.vrabec@upb.de}

\begin{abstract}

A new version release (2.0) of the molecular simulation tool {\it ms}2 [S. Deublein et al., Comput. Phys. Commun. 182 (2011) 2350] is presented. Version 2.0 of {\it ms}2 features a hybrid parallelization based on MPI and OpenMP for molecular dynamics simulation to achieve higher scalability. Furthermore, the formalism by Lustig [R. Lustig, Mol. Phys. 110 (2012) 3041] is implemented, allowing for a systematic sampling of Massieu potential derivatives in a single simulation run. Moreover, the Green-Kubo formalism is extended for the sampling of the electric conductivity and the residence time. To remove the restriction of the preceding version to electro-neutral molecules, Ewald summation is implemented to consider ionic long range interactions. Finally, the sampling of the radial distribution function is added.  

\end{abstract}

\maketitle

\section{Introduction}
\label{intro}
Molecular modeling and simulation is a technology central to many areas of research in academia and industry. With the advance of computing power, the scope of application scenarios for molecular simulation is widening, both in terms of complexity of a given simulation and in terms of high throughput. Nowadays, e.g. the predictive simulation of entire phase equilibrium diagrams has become feasible. However, in order to rely on simulation results, the methodology needs to be sound and the implementation must be thoroughly verified. In its first release ~\cite{ms2}, we have introduced the molecular simulation tool $ms$2. Results from $ms$2 have been verified and the implementation was found to be robust and efficient.

As described in Section \ref{hybrid}, in Version 2.0 of the simulation tool $ms$2 the existing molecular dynamics (MD) MPI parallelization was hybridized with OpenMP, leading to an improved performance on multi-core processors. Furthermore, the new release offers a wider scope of accessible properties. In particular, $ms$2 was extended to calculate Massieu potential derivatives in a systematic manner, cf. section \ref{massieu}. This augments the range of sampled properties significantly and, as was demonstrated in ~\cite{Rut13}, it allows to straightforwardly develop competitive fundamental equations of state from a combination of experimental VLE data and molecular simulation results. Lastly, besides being now capable of simulating ionic substances, the time and memory demand for calculating transport properties was reduced significantly (section \ref{algo}).

$ms$2 is freely available as an open source code for academic users at \url{www.ms-2.de}.

\section{Hybrid MPI \& OpenMP Parallelization}
\label{hybrid}
The molecular simulation tool $ms$2 focuses on thermodynamic properties of homogeneous fluids. Therefore, 
systems investigated with $ms$2 typically contain on the order of $10^3$ molecules. While for Monte Carlo simulations a perfect scaling behavior up to large numbers of cores can be trivially achieved, MD domain decomposition -- the de facto standard for highly scalable MD -- is not feasible for such system sizes, because the cut-off radius is in the same range as half the edge length of the simulation volume. This excludes domain decomposition and limits the scalability of the MPI parallelization. The present release of $ms$2 features an OpenMP parallelization, which was hybridized with MPI. At the point where MPI communication becomes a bottleneck, a single process still has enough load to distribute to multiple threads, improving scalability.

Three parts of $ms$2 were parallelized with OpenMP: the interaction partner search, the energy and the force calculations. All OpenMP parallel regions rely on loop parallelism, as the compute intensive parts of the algorithm all feature a loop over the molecules. In the force calculation, race conditions need to be considered, because every calculated force is written to both interacting molecules. Introducing atomic updates or critical sections leads to massive overheads. Instead, it is more efficient to assign forces from individual interactions to the elements of a list (or an array) which is subsequently summed up. The same holds true for torques.

\begin{figure}[h]
\centering
\includegraphics[width=8.333cm]{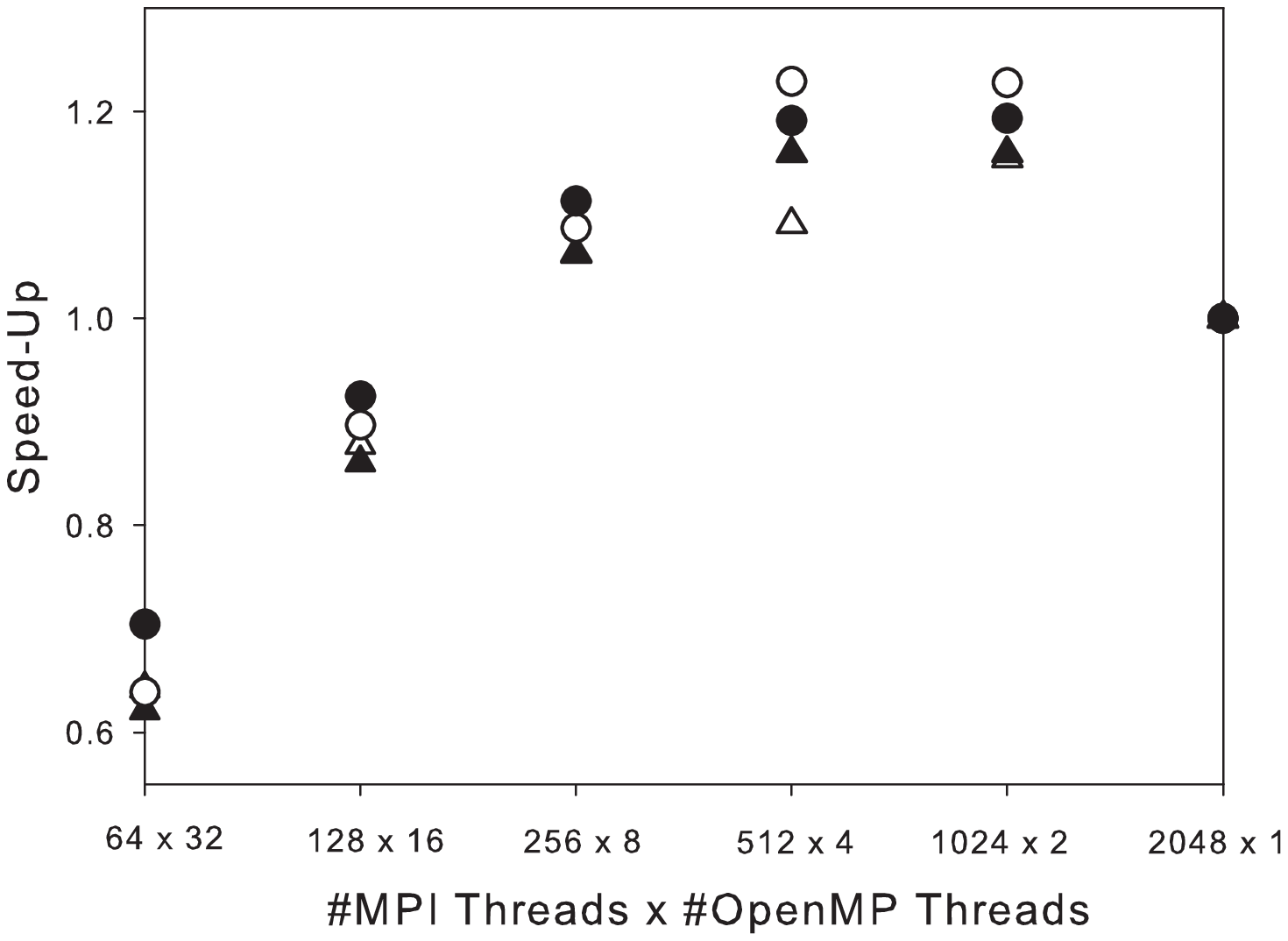}
\caption{Speed-up of hybrid MPI/OpenMP vs. pure MPI for 2048 cores, varying number of threads per MPI process and 8192 molecules (solid circles), 4096 molecules (empty circles), 2048 molecules (solid triangles), 1024 molecules (empty triangles)}
\label{Hybrid}
\end{figure}

In Figure~\ref{Hybrid} the speed-up of hybrid MPI/OpenMP vs. pure MPI is plotted for 2'048 cores, varying the number of threads per MPI process and the number of molecules in the simulation volume. As can be seen, using 2 to 4 threads per MPI process delivers a speed-up of around 20\% for 2'048 cores. The evaluation of the hybrid parallelization algorithm was performed on a CRAY XE6 Supercomputer at the High Performance Computing Center in Stuttgart, which has an overall peak performance of one PFLOPS. It consists of 3552 nodes, each equipped with two AMD Opteron 6276 (Interlagos) processors. Each processor has 16 cores, sharing eight FPUs (Floating Point Units). Nodes are equipped with 32 GB RAM and are interconnected by a high-speed CRAY Gemini network. Additional runtime performance comparisons with the simulation tool \textit{GROMACS}~\cite{gromacs} are listed in Table \ref{tab1}.

\begin{table}
  \centering
  \caption{Runtime performance results with $ms$2 release 2.0 and GROMACS v4.6.5~\cite{gromacs} for MD simulations with pure water at 298.15 K and  55.345 $\text{mol} \cdot \text{dm}^{-3}$. The number of time steps were 100 000 for every simulation, the cutoff radius was identical for simulations with the same number of particles. All simulations were performed on the same computer cluster.}
    \begin{tabular}{cccccc}
    cores & threads & $N$ & gromacs / s & $ms$2(RF) / s & $ms$2(EW) / s\\
    \hline
8	&	8 MPI 	&	500	&	164	&	416	&	785	\\
8	&	8 MPI 	&	1000	&	299	&	874	&	1607	\\
8	&	8 MPI 	&	2000	&	1284	&	4461	&	6777	\\
16	&	16 MPI 	&	500	&	95	&	233	&	415	\\
16	&	16 MPI 	&	1000	&	166	&	477	&	848	\\
16	&	16 MPI 	&	2000	&	678	&	2298	&	3506	\\
32	&	32 MPI 	&	500	&	62	&	152	&	245	\\
32	&	32 MPI 	&	1000	&	106	&	296	&	487	\\
32	&	32 MPI 	&	2000	&	361	&	1286	&	1898	\\
64	&	64 MPI 	&	500	&	40	&	119	&	166	\\
64	&	64 MPI 	&	1000	&	65	&	228	&	324	\\
64	&	64 MPI 	&	2000	&	220	&	814	&	1261	\\
128	&	128 MPI 	&	500	&	38	&	105	&	131	\\
128	&	128 MPI 	&	1000	&	51	&	197	&	247	\\
128	&	128 MPI 	&	2000	&	147	&	557	&	727	\\
\hline
8	&	1 MPI, 8 OMP/MPI	&	500	&	167	&	483	&		\\
8	&	1 MPI, 8 OMP / MPI	&	1000	&	323	&	975	&		\\
8	&	1 MPI, 8 OMP / MPI	&	2000	&	1416	&	4831	&		\\
16	&	2 MPI, 8 OMP / MPI	&	500	&	105	&	253	&		\\
16	&	2 MPI, 8 OMP / MPI	&	1000	&	186	&	517	&		\\
16	&	2 MPI, 8 OMP / MPI	&	2000	&	763	&	2514	&		\\
32	&	4 MPI, 8 OMP / MPI	&	500	&	75	&	167	&		\\
32	&	4 MPI, 8 OMP / MPI	&	1000	&	121	&	316	&		\\
32	&	4 MPI, 8 OMP / MPI	&	2000	&	418	&	1362	&		\\
64	&	8 MPI, 8 OMP / MPI	&	500	&	60	&	119	&		\\
64	&	8 MPI, 8 OMP / MPI	&	1000	&	92	&	217	&		\\
64	&	8 MPI, 8 OMP / MPI	&	2000	&	261	&	785	&		\\
128	&	16 MPI, 8 OMP / MPI	&	500	&	49	&	101	&		\\
128	&	16 MPI, 8 OMP / MPI	&	1000	&	74	&	172	&		\\
128	&	16 MPI, 8 OMP / MPI	&	2000	&	170	&	496	&		\\
    \hline
    \multicolumn{6}{l}{($N$) Number of water molecules}\\
    \multicolumn{6}{l}{(RF) simulations with reaction field correction.}\\
    \multicolumn{6}{l}{(EW) simulations with Ewald summation.}\\
    \end{tabular}
    \label{tab1}
\end{table}

\section{Massieu potential derivatives}
\label{massieu}
$ms$2 version 2.0 features evaluating free energy derivatives in a systematic manner, thus greatly extending the thermodynamic property types that can be sampled in single simulation runs. The approach is based on the fact that the fundamental equation of state contains the complete thermodynamic information about a system, which can be expressed in terms of various thermodynamic potentials~\cite{MunsterBook}, e.g. internal energy $E(N,V,S)$, enthalpy $H(N,p,S)$, Helmholtz free energy $F(N,V,T)$ or Gibbs free energy $G(N,p,T)$, with number of particles $N$, volume $V$, pressure $p$, temperature $T$ and entropy $S$. These representations are equivalent in the sense that any other thermodynamic property is essentially a combination of derivatives of the chosen form with respect to its independent variables. The form $F/T(N,V,1/T)$, known as the Massieu potential, is preferred in molecular simulations due to practical reasons ~\cite{Lus11, Lus12}. The statistical mechanical formalism of Lustig allows for the simultaneous sampling of any $A^r_{mn}$ in a single $NVT$ ensemble simulation for a given state point ~\cite{Lus11, Lus12, Lus94a, Lus94b}
\begin{eqnarray} 
\frac{\partial^{m+n}(F/(RT))}{\partial\beta^{m} \partial \rho^{n}} \beta^m \rho^n  \equiv A_{mn} = A^i_{mn} + A^r_{mn}\mbox{   ,}
\label{eq:deriv}
\end{eqnarray}
where $R$ is the gas constant, $\beta \equiv 1/T$ and $\rho \equiv N/V$. $A_{mn}$ can be separated into an ideal part $A^i_{mn}$ and a residual part $A^r_{mn}$ ~\cite{RowBook}. The calculation of the residual part is the target of molecular simulation and the derivatives $A^r_{10}$, $A^r_{01}$, $A^r_{20}$, $A^r_{11}$, $A^r_{02}$, $A^r_{30}$, $A^r_{21}$ and $A^r_{12}$ were implemented in $ms$2 for $NVT$ ensemble simulations. The ideal part can be obtained by independent methods, e.g. from spectroscopic data or ab initio calculations. However, it can be shown that for any $A_{mn}=A^i_{mn}+A^r_{mn}$, where $n>0$, the ideal part is either zero or depends exclusively on the density, thus it is known by default ~\cite{Lus12}. Note that the calculation of $A^r_{00}$ still requires additional concepts such as thermodynamic integration or particle insertion methods. From the first five derivatives $A_{10}$, $A_{01}$, $A_{20}$, $A_{11}$, $A_{02}$ every measurable thermodynamic property can be expressed (see the supplementary material for a list of properties) with the exception of phase equilibria. A detailed description of the implementation is in the supplementary material, here, only an overview is provided.\\  
The calculation of the derivatives up to the order of $n=2$ requires the explicit mathematical expression of ${\partial U}/{\partial V}$ and ${\partial^2 U}/{\partial V^2}$ with respect to the applied molecular interaction pair potential and has to be determined analytically beforehand ~\cite{Lus11, Lus12}. The general formula for ${\partial^n U}/{\partial V^n}$ can be found in Ref. ~\cite{Lus94b}. For common molecular interaction pair potentials, like the Lennard-Jones potential~\cite{allen1987, frenkel2003}, describing repulsive and dispersive interactions, or Coulomb's law, describing electrostatic interactions between point charges, the analytical formulas for ${\partial U}/{\partial V}$ and ${\partial^2 U}/{\partial V^2}$ can be obtained straightforwardly.\\
As molecular simulation is currently limited to operate with considerably fewer particles than real systems, the effect of the small system size thus has to be counter-balanced with a contribution to $U$ and ${\partial^n U}/{\partial V^n}$ called long range correction (LRC) ~\cite{allen1987, frenkel2003}. The mathematical form of the LRC depends on the molecular interaction potential and the cut-off method (site-site or center-of-mass cut-off mode) applied. For the Lennard-Jones potential, the LRC scheme was well described in the literature for both the site-site~\cite{Lus11, Mei06} and the center of mass cut-off mode~\cite{Lus94b, Lus88}. The reaction field method~\cite{RFmethod} was the default choice in the preceding version of $ms$2 for the LRC of electrostatic interactions modelled by considering charge distributions on molecules. The usual implementation of the reaction field method combines the explicit and the LRC part in a single pair potential~\cite{RFmethod, siteRFmethod} from which ${\partial^n U}/{\partial V^n}$ (including the LRC contribution) is directly obtainable. However, practical applications show that the electrostatic LRC of ${\partial U}/{\partial V}$ and ${\partial^2 U}/{\partial V^2}$ can be neglected in case of systems for which the reaction field method is an appropriate choice. E.g., the contribution of the electrostatic LRC for a liquid system ($T=298$ K and $\rho= 45.86$ mol/l) consisting of only 200 water and 50 methanol molecules with a very short cut-off radius of $20\%$ of the edge length of the simulation volume is still $<<1\%$ for both ${\partial U}/{\partial V}$ and ${\partial^2 U}/{\partial V^2}$. The supplementary material contains detailed elaborations on the LRC for the Lennard-Jones potential.

\section{Algorithmic Developments}
\label{algo}
\paragraph{Transport property calculations}
In $ms$2, transport properties are determined via equilibrium MD simulations by means of the Green-Kubo formalism~\cite{gubbins1972}. This formalism offers a direct relationship between transport coefficients and the time integral of the autocorrelation function of the corresponding fluxes. An extended time step was defined for the calculation of the fluxes, the autocorrelation functions and their integrals. The extended time step is $n$ times longer than the specified MD time step, where $n$ is a user defined variable. The autocorrelation functions are hence evaluated in every $n$-th MD time step. As a consequence, the memory demand for the autocorrelation functions was reduced and the restart files, which contain the current state of the autocorrelation functions and time integrals, become accordingly smaller. In addition, the overall computing time of the MD simulation was reduced significantly.\\ 

\paragraph{Ewald summation}
Ewald summation~\cite{allen1987, frenkel2003} was implemented for the calculation of electrostatic interactions between point charges. It extends the applicability of $ms$2 to thermodynamic properties of e.g. ions in solutions. In Ewald summation, the electrostatic interactions according to Coulomb's law are divided into two contributions: short-range and long-range. The short-range term includes all charge-charge interactions at distances smaller than the cut-off radius. The remaining contribution is calculated in Fourier space and only the final value is transformed back into real space. This allows for an efficient calculation of the long-range interactions between the charges. The algorithm is well described in literature. Currently, some of the new features, the calculation of Massieu potential derivatives and Hybrid MPI \& OpenMP Parallelization for MD, are not available together with Ewald summation.

\section{Property Calculations}
\label{thermoprop}

\paragraph{Radial distribution function}
The radial distribution function (RDF) $g(r)$ is a measure for the microscopic structure of matter. It is defined by the local number density around a given position within a molecule $\rho^L(r)$ in relation to the overall number density $\rho = N/V$

\begin{equation} g(r) = \dfrac{\rho^L(r)}{\rho} = \dfrac{1}{\rho}\dfrac{\mathrm{d}N(r)}{\mathrm{d}V} = \dfrac{1}{4 \pi r^2 \rho} \dfrac{\mathrm{d}N(r)}{\mathrm{d}r}. \label{Eq:gr} \end{equation}
Therein, $\mathrm{d}N(r)$ is the differential number of molecules in a spherical shell volume element $\mathrm{d}V$, which has the width $\mathrm{d}r$ and is located at the distance $r$ from the regarded position. $g(r)$ can be evaluated for every molecule of a given species. 

In the present release of $ms$2, the RDF can be calculated during MD simulation runs for pure components and mixtures on the fly. The RDF is sampled between all LJ sites. In order to evaluate RDFs for arbitrary positions, say point charge sites, superimposed dummy LJ sites with the parameters $\sigma = \epsilon = 0$ have to be introduced in the potential model file by the user.

\paragraph{Electric conductivity}
\noindent
The evaluation of the electric conductivity $\sigma$ was implemented in $ms$2 version 2.0, being a measure for the flow of ions in solution. The Green-Kubo formalism~\cite{gubbins1972} offers a direct relationship between $\sigma$ and the time-autocorrelation function of the electric current flux $\bm{j_e}(t)$ ~\cite{HansenBook}
\begin{equation}
\sigma=\frac{1}{3Vk_{\mathrm{B}}T}\int_0^{\infty}~\big\langle \bm{j_e}(t)\cdot \bm{j_e}(0) \big\rangle  \mathrm{d} t \mbox{   ,}
\label{Erdalkali_elcond}
\end{equation}
where $k_\texttt{B}$ is Boltzmann’s constant. The electric current flux is defined by the charge $q_k$ of ion $k$ and its velocity vector $\bm{v}_k$ according to
\begin{equation}
\bm{j_e}(t)=\sum_{k=1}^{N_{j}} q_k \cdot \bm{v}_{k}(t) \mbox{   ,}
\label{elflux}
\end{equation}
where $N_{j}$ is the number of molecules of component $j$ in solution. Note that only the ions in the solution have to be considered, not the electro-neutral molecules.
For better statistics, $\sigma$ is sampled over all independent spatial elements of $\bm{j_e}(t)$.

\paragraph{Thermal conductivity of mixtures}
\noindent
In the previous version of $ms2$ the determination of the thermal conductivity by means of the Green-Kubo formalism was implemented for pure substances only. In the present release, the calculation of the thermal conductivity was extended to multi-component mixtures. The thermal conductivity $\lambda$ is given by the autocorrelation function of the elements of the microscopic heat flow ${J}^{x}_{q}$
\begin{equation}\label{themalcond}
\lambda=\frac{1}{Vk_BT^{2}}\int_{0}^{\infty} dt~\big\langle {J}^x_q(t)\cdot {J}^x_q(0)\big\rangle.
\end{equation}
\noindent In mixtures, energy transport and diffusion occur in a coupled manner, thus, the heat flow for a mixture of $n$ components is given by~\cite{evans1}
\begin{eqnarray}
\mathbf{J}_q &=& \frac{1}{2} \sum_{i=1}^{n} \sum_{k=1}^{N_i}\left[ m_{i}^{k}\left({v}_{i}^{k}\right)^2+ \nonumber
\mathbf{w}_i^k \mathbf{I}_i^k \mathbf{w}_i^k+\sum_{j=1}^{n}\sum_{l \neq k}^{N_j} u\left(r_{ij}^{kl}\right)\right]\cdot \mathbf{v}_i^k \\
             & -&\frac{1}{2}\sum_{i=1}^{n} \sum_{j=1}^{n} \sum_{k=1}^{N_i} \sum_{l \neq k}^{N_j} \bm{r}_{ij}^{kl}\cdot
\big( \mathbf{v}_i^k \cdot \frac{\partial u\left(r_{ij}^{kl}\right)}{\partial \bm{r}_{ij}^{kl}}+\mathbf{w}_i^k \mathbf{\Gamma}_{ij}^{kl}\big)- \sum_{i=1}^{n} h_{i}\sum_{k=1}^{N_i} \mathbf{v}_{i}^{k}\mbox{   ,}\label{heatflow} 
\end{eqnarray}
\noindent where $\mathbf{w}^k_i$ is the angular velocity vector of molecule $k$ of component $i$ and $\mathbf{I}^k_i$ its matrix of angular momentum of inertia. $u\left(r_{ij}^{kl}\right)$ is the intermolecular potential energy and $\mathbf{\Gamma}_{ij}^{kl}$ is the torque due to the interaction of molecules $k$ and $l$. The indices $i$ and $j$ denote the components of the mixture. $h_{i}$ is the partial molar enthalpy. It has to be specified as an input in the $ms2$ parameter file and can be calculated from $NpT$ simulations.

\paragraph{Residence time}
The residence time $\tau_{j}$ defines the average time span that a molecule of component $j$ remains within a given distance $r_{ij}$ around a specific molecule $i$. It is given by the autocorrelation function
\begin{equation}
\tau_{j} = \int_{t=0}^{\infty} \left\langle \frac{1}{n_{ij}(0)} \sum_{k=1}^{n_{ij}(0)}\Theta_k(t)\Theta_k(0)  \right\rangle \mathrm{d}t  \mbox{   ,}
\end{equation}
where $t$ is the time, $n_{ij}(0)$ the solvation number around molecule $i$ at $t=0$ and $\Theta$ is the Heaviside function, which yields unity, if the two molecules are within the given distance, and zero otherwise. 
Following the proposal of Impey et al.~\cite{Impey1983}, the residence time explicitly allows for short time periods during which the distance between the two molecules exceeds $r_{ij}$. Also, the solvation number $n_{ij}$ can be evaluated on the fly 

\begin{equation}
 n_{ij} = 4\pi \rho_{j} \int_0^{r_{\mathrm{min}}} r^2g_{ij}(r) \mathrm{d}r \mbox{   ,}
\end{equation}
where $\rho_{j}$ is the number density of component $j$ and $r_{\mathrm{min}}$ is the distance up to which the solvation number is calculated. 

\noindent

\section*{Acknowledgments}
The authors gratefully acknowledge financial support by the
BMBF "01IH13005A SkaSim: Skalierbare HPC-Software f\"ur molekulare Simulationen in der chemischen Industrie" and computational support by the High Performance Computing Center Stuttgart (HLRS) under the grant MMHBF2. The present research was conducted under the auspices of the Boltzmann-Zuse Society for
Computational Molecular Engineering (BZS).

\providecommand{\noopsort}[1]{}\providecommand{\singleletter}[1]{#1}%

\end{document}